\documentstyle[11pt,paspconf,epsf]{article}

\newcommand {\chem}[2] {$\rm{}^{#2}\kern-0.8pt#1$}
\newcommand {\mass}[1] {$\rm #1\,M_{\sun}$}

\newcommand {\reac}[6] {$\rm\,{}^{#2}\kern-0.8pt{#1}\,({#3}\,,{#4})
                        \,{}^{#6}\kern-0.8pt{#5}\,$}
\def\dot#1{\hbox{$#1$ \kern -1.8ex \raisebox{1.7ex}.}\,}

\begin{document}
\title{Yields from Low- and Intermediate-Mass Stars
}
{\footnote{Workshop {\sl Abundance profiles: Diagnostic Tools for Galaxy 
History}, 1997, Qu\'ebec, ASP Conference Series}}
\author{Corinne Charbonnel}
\affil{Laboratoire d'Astrophysique de Toulouse, CNRS UMR 5572, \\
14, av. E.Belin, F-31400 Toulouse, France} 
\begin{abstract}
Stellar nucleosynthesis is the corner-stone of many astrophysical problems.
Its understanding, which can be tested by countless observations, 
leads to insights into the stellar structure and evolution, 
and provides crucial clues to the physics of the galaxies and of the universe. 
Precise answers can be given to the questions 
``When, where and how the chemical elements are synthesized in stars?",
and will be summarized in this paper for what concerns stars of low and 
intermediate mass. 
However, in spite of the observational confirmation of many 
predictions, important and complementary data reveal some weaknesses in the 
theory of stellar physics. In particular, the mixing processes of chemicals 
inside the stars, as well as the mass loss and ejection mechanisms, are 
poorly known. 
Their rudimentary treatment imposes the use of parameters which strongly 
influence the predicted chemical yields.  
In our discussion, we will underline the future developments that may lead 
to quantitative changes in the predictions of chemicals production by low and 
intermediate mass stars.  
\end{abstract}
\keywords{Keyword1 -- Keyword2 -- Etc}


\section{Evolution and Nucleosynthesis up to the AGB Phase}
We restrict our discussion to the case of single low and intermediate mass 
stars (LIMS), which are those that develop an electron-degenerate 
carbon-oxygen core after the center He-exhaustion and do not proceed through 
the carbon and heavier elements burning phases 
(The case of massive stars is treated in this volume by N.~Prantzos). 
This definition places an upper limit of 6--8\,M$_{\odot}$ 
(see Maeder \& Meynet 
1989 for a discussion on this value which depends mainly on the mixing 
prescription and mass loss rates used in the evolutionary models). 

\subsection{The Main Trends}
The nucleosynthesis occuring during the evolution of LIM stars is well 
known.
For what concerns the stellar surface abundances and the chemical enrichment 
of the interstellar matter, the only interesting nucleosynthesis process 
in these stars up to the AGB phase is hydrogen burning (the products of
central helium burning remain indeed trapped into the white dwarf).

Nuclear reactions involving elements from D to C begin to occur on the 
pre-main sequence 
for temperatures higher than 10$^6$\,K. During this phase, D burns to $^3$He;
$^6$Li and $^7$Li are destroyed by p-capture, and their abundances decrease
at the surface of stars with masses lower than $\sim$ 1\,M$_{\odot}$.

On the main sequence (MS), hydrogen burning occurs in a radiative core and is 
dominated by the pp-reactions in stars with masses lower than $\sim$ 
1.1\,M$_{\odot}$, while the CNO bi-cycle dominates in the convective core of 
more massive stars. 
Partial pp-chain burning builds up a peak of $^3$He. 
Slightly further in, $^{12}$C and $^{16}$O are partially converted into 
$^{14}$N.
A peak in the $^{13}$C abundance develops. 
Below, $^{18}$O is burned to $^{15}$N, and $^{17}$O is slightly enhanced 
from partial burning of $^{16}$O (a peak of $^{17}$O is present).
The chemical profiles 
at the end of the MS depend both on mass and metallicity.  

After the central H exhaustion, nuclear energy is released by the
hydrogen burning shell (HBS) that surrounds the He-core which
becomes electron degenerate in low mass stars (LMS, i.e., with masses 
lower than 2-2.3M\,$_{\odot}$). The star expands and the convective 
envelope penetrates deeply inwards until $\sim 70 \%$ of the total mass 
is convective, 
reaching the region where central H-burning occured. 
The first dredge-up (1Dup) leads to the first
important modification of the surface abundances.

While the star is on the horizontal branch, the convective core where
helium burns is surrounded by a thin HBS.
Following core helium exhaustion, the structure readjusts to shell
burning (HeBS, HBS), and the star strongly expands while it enters the 
early-AGB phase. The envelope penetrates in the deep layers. 
In stars with masses higher than $\sim$ 4M\,$_{\odot}$, a second dredge-up 
(2Dup) mixes to the surface the products of complete H-burning (contrary to
the 1Dup, the HBS extinguishes). 
Soon after, the first thermal pulse (TP) occurs (see \S 2.1).

\subsection{The First and Second Dredge-Up. Theoretical Predictions}

During the 1 and 2Dup, convective mixing and induced dilution 
modify the surface abundances of the elements altered by H-burning. 
The extent of these changes depends both on stellar mass and metallicity (Z): 
For stars with masses higher than 2-2.3M\,$_{\odot}$, the 1Dup is almost 
negligible at low metallicity, and its importance increases with Z. 
The effect of the 2Dup in lower mass stars is negligible whatever the 
metallicity. 
The predictions obtained recently by different groups agree quantitatively 
well (Schaller et al. 1992; Dearborn 1992; Bressan et al. 1993; El Eid 1994;
Forestini \& Charbonnel 1997 (hereafter FC97); Boothroyd \& Sackman 1997 
for the most recent papers), except for a few features. 
Basically, one can summarize as follows: 

In LIM stars, the surface mass fraction of $^3$He increases by factors 2
to 6. Once in the convective layers of the red giant, $^3$He is preserved 
because of the too cool temperature in these regions. 
Since $^7$Li is burned on the main sequence in the
regions where the temperature is higher than about $2.5 \cdot 10^6$\,K, the 
surface abundance of this element strongly decreases after the 1Dup.
The total carbon abundance decreases by approximately 30$\%$; there
is no carbon depletion for $<$1M\,$_{\odot}$ stars, and less than a factor 
of 2 at higher masses. 
From an initial value of the order of 90, the carbon isotopic ratio
decreases down to about 20-30 due to the 1Dup; since after central HeB, 
almost all the CNO isotopes have been converted into $^{14}$N, the 2Dup has 
little effect on the $^{12}$C/$^{13}$C ratio. 
The $^{14}$N abundance increases by about 80$\%$ and the
$^{12}$C/$^{14}$N decreases. 
The $^{16}$O/$^{17}$O ratio does not change for stars with masses lower
than 1M\,$_{\odot}$, but it decreases when the convective envelope reaches 
partially the $^{17}$O pocket (in stars with masses between 1 and 
2M\,$_{\odot}$) or completely engulfes it (in more massive stars; there, 
uncertainties in the $^{17}$O-destruction rates affect the predictions 
for the final $^{16}$O/$^{17}$O ratio). 
The very slight rise of the $^{16}$O/$^{18}$O ratio is an increasing 
function of the stellar mass. 
The total abundance of oxygen and of all heavier elements is not affected 
by the 1 and 2Dup. 

\subsection{Comparisons with the Observations}
Many observations exist to which one can confront the predictions
described above. 
For what concerns stars with masses higher than about 2M\,$_{\odot}$, no major
conflict appears. 
The theoretical post dredge-up values of the carbon isotopic ratio are 
slightly lower but in agreement with the observations in galactic cluster 
giants (Gilroy 1989). Red giants present \chem{O}{16}/\chem{O}{17} 
ratios between 300 to 1000 and \chem{O}{16}/\chem{O}{18} ratios in the 
range 400 to 600 (Harris et al. 1988; Smith \& Lambert 1990a), as predicted.

The case of lower mass stars is however problematic (see Charbonnel et
al. 1998, hereafter CBW98, for references). 
In most of the LM evolved stars, the observed conversion of $^{12}$C to
$^{13}$C and $^{14}$N greatly exceeds the levels expected from standard 
stellar models. 
The $\rm ^{12}C/^{13}C$ ratio even reaches the near-equilibrium value
in many Population II RGB stars.  
This problem also occurs, but to a somewhat lower extent,
in evolved stars belonging to open clusters with turnoff masses
lower than 2\,M$_{\odot}$.
In halo giants, the lithium abundance continues to decrease
after the completion of the first dredge-up.
A continuous decline in carbon abundance with increasing stellar
luminosity along the RGB is observed in globular clusters such as
M92,
M3 and M13,
M15,
NGC 6397,
NGC 6752 and M4.
In some globular clusters (M92, 
M15, 
M13, 
$\omega$ Cen), 
giants exhibit evidence in their atmospheres for O$\rightarrow$N
processed material.
In addition to the O versus N anticorrelation, the existence of Na
and Al versus N correlations and Na and Al versus O anticorrelations in
a large number of globular cluster red giants has been clearly confirmed.

\subsection{Towards More Complete Stellar Models. 
Consequences of an Extra-Mixing on the RGB}

None of the behaviours described above is predicted by the standard 
stellar theory.
These observations suggest that, while they evolve on the RGB,
LM stars undergo an extra-mixing in the region situated
between the hydrogen burning shell (where the material is processed
through the CN-cycle and possibly the ON-cycle) and the deep convective
envelope.
This extra-mixing, which appears to be efficient only after the
luminosity-function bump (Gilroy \& Brown 1991; Charbonnel 1994; CBW98),
adds to the standard first dredge-up to modify the surface abundances. 
Recently, different groups have simulated an extra-mixing process 
in order to reproduce the CNO abundances in RGB stars
(Denissenkov \& Weiss 1995; Wasserburg et al. 1995; Boothroyd \& Sackman 1997). 
Some authors attempted to relate the
extra-mixing with physical processes, among which rotation and mass loss 
seems to be the most promising (Sweigart \& Mengel 1979; Charbonnel 1995). 

In any case, these ``non-standard" models (related or not to a physical 
process), make a common crucial prediction : The mechanism(s) which is (are) 
responsible for the chemical anomalies on the RGB must lead to the 
destruction of $^3$He by a large factor in the bulk of the envelope 
material (Hogan 1995; Charbonnel 1995). This result strongly modifies 
the actual contribution of LM stars to the galactic evolution of this
element (see Tosi 1996; Charbonnel 1997). It prevents its overproduction 
and helps understanding the recent measurements of $^3$He in the local 
interstellar cloud (Gloeckler \& Geiss 1996), in galactic H\,{\sc ii} regions 
(Balser et al. 1994) and in planetary nebulae (Rood et al. 1992; 
Balser et al. 1997).

The physics of the extra-mixing process in LMS has to be better understood. 
Detailed simulations, with a consistent treatment of the transport of 
matter and angular momentum, have to be carried out for different stellar 
masses and metallicities, and various mass loss and rotation histories. 
The impact of this process on the behavior of various elements in RGB stars 
(C $\searrow$, Na $\nearrow$, O $\searrow$, Al $\nearrow$, Mg O
$\searrow$) 
and on the precise yields of $^3$He has to be investigated
in details. Consequences for the energy production in the HBS and for 
the HB morphology may not be neglected (Sweigart 1997). 

\section{Evolution and Nucleosynthesis on the Thermally-Pulsing
AGB phase (TP-AGB)}

\subsection{Structural Evolution}
The structure of an AGB star is characterized by an electron-degenerate
CO core of mass between 0.5 and 1.2\,M$_{\odot}$ which will become the 
white dwarf remnant, surrounded by an HeBS which is thermally unstable
and an HBS. This very small and compact region is confined inside a deep
and very extended convective envelope. 
During this phase, the evolution of the star is dominated by the recurrent 
thermal pulses and by a very important mass loss. 
%
After each pulse extinction, the convective envelope deepens. 
Eventually, the third dredge-up (3Dup), which is a repeating phenomenon, 
occurs. 
Contrary to the 1 and 2Dup, the 3Dup successively mixes up to the 
surface {\it (i)} material that has experienced H-burning in the thin 
HBS but also {\it (ii)} part of the region where the thermal pulse 
nucleosynthesis operated just before. 
As we shall see now, both regions give rise to very different -- and sometimes 
opposite -- chemical pollutions of the convective envelope. 
Nuclear reactions occuring inside the convective envelope may modify further 
the surface abundances in IMS. 
The newly synthetized nuclides are ejected into the interstellar
medium (ISM) through the strong mass loss and the planetary nebulae ejection.  
The TP-AGB stars play thus a crucial role for the chemical
evolution of galaxies, especially for some elements like \chem{He}{3},
\chem{Li}{7}, \chem{F}{19}, \chem{Al}{26}, and nuclides heavier than
\chem{Fe}{56}. 

\subsection{Nucleosynthesis on the TP-AGB}
Four different nucleosynthesis sites can be distinguished.

\noindent (1) The HBS during the interpulse. 
~~In this region, most of the nuclear energy is released by the CNO bi-cycle.
Whatever the total mass of the AGB star, 
\chem{C}{12}, \chem{O}{17}, \chem{O}{18} and \chem{F}{19} are significantly 
destroyed in the HBS, while \chem{N}{14} and \chem{N}{15} are produced.
The abundance of \chem{C}{13} is slightly increased, except in the 
more massive AGB stars. 
The rather high temperature inside the HBS also allows the activation of the
NeNa (\chem{Ne}{22} and \chem{Na}{23} are produced in $\le$ \mass{5} stars 
and slightly destroyed in more massive objects) and MgAl chains. 
This in turn leads to an important production of \chem{Al}{26}.
However, as the HBS mean temperature increases with time along the 
TP-AGB, \chem{Al}{26} begins to be partially destroyed by 
$\rm{(p , \gamma)}$ reactions later on (Forestini et al. 1991; Guelin et 
al. 1995).

\noindent (2) The HeBS. 
~~Through the (3$\alpha$) reaction, the HeBS mainly produces 
\chem{C}{12} and, to a lower extent, \chem{O}{16}. 
During the TP, a convective tongue develops and penetrates the intershell 
region, bringing the \chem{C}{12} close to the HBS and ingesting the HBS ashes.
Of crucial importance is the injection of \chem{C}{13} which immediately 
burns through the \reac{C}{13}{\alpha}{n}{}{} reaction. 
The produced neutrons are captured through $\rm{(n , \gamma)}$ and 
$\rm{(n , p)}$ reactions. 
The simultaneous presence, inside a mixed He-burning region, of neutrons
and protons leads to a very specific and unique nucleosynthesis.
In particular, \chem{F}{19} can be produced through the reaction pathway 
$^{18}$O(p,$\alpha)^{15}$N($\alpha,\gamma)^{19}$F (Forestini et al. 1992).
However, the more massive the AGB star, the less efficient the
\chem{F}{19} production. Due to HBB, the most massive AGB stars even
present negative net yields for that element. LM-AGB stars are probably
the most efficient \chem{F}{19} producers in galaxies (Mowlavi et al.
1996; FC97). 
In addition, elements heavier than iron are produced in this region via 
the classical s-process nucleosynthesis. 
However, as will be discussed in \S2.4, the amount of \chem{C}{13} spread out 
by the HBS is by far (i.e. at least a factor of 10 or more) insufficient to 
explain quantitatively the s-process (Sackmann \& Boothroyd 1991). 

\noindent (3) The intershell region. 
~~Such low quantities of \chem{C}{13} ingested 
by thermal pulses are partly explained by the partial \chem{C}{13} radiative 
burning, operating through the \reac{C}{13}{\alpha}{n}{O}{16} reaction 
at the bottom of the intershell region during the interpulse phase 
(Forestini 1991). 
This third  burning shell could at least partially explain the production 
of the s-nuclides in a radiative zone. This was shown by Straniero 
et al. (1995) who had to artificially increase the amount of \chem{C}{13} 
in the intershell in order to study the s-process nucleosynthesis.

\noindent (4) The base of the convective envelope. 
~~If the temperature T$_{cb}$ at the base of the deep convective envelope
is high enough (i.e. $20 \cdot 10^6$\,K),
proton burning occurs inside the envelope.
This so-called hot-bottom burning (HBB, Sackmann et al. 1974)
occurs in stars more massive than $\sim$ 4\,M$_{\odot}$. 
As mixing up to the cool surface is very efficient, the effective nuclear 
reaction rates are lower than in radiative H-burning, leading to specific 
nucleosynthesis signatures. 
For T$_{cb} > 40 \cdot 10^6$\,K, 
\chem{Li}{7} can be produced from \chem{He}{3}
via the Cameron \& Fowler (1971) scenario (Sackman \& Boothroyd 1992;
FC97; see N.Mowlavi's contribution in the present volume).
For T$_{cb} > 60 \cdot 10^6$\,K, the CN cycle begins to operate
partially, i.e. \chem{C}{12} is converted to \chem{C}{13} that
accumulates. As first suggested by Wood et al. (1983), this conversion can 
efficiently reduce the \chem{C}{12}/\chem{O}{16} ratio and possibly, if HBB 
is strong enough, this can prevent the formation of C stars.
If T$_{cb}$ $>$ $80 \cdot 10^6$\,K typically, the CN cycle almost operates
at equilibrium through the whole envelope, leading to a
\chem{C}{12}/\chem{C}{13} isotopic ratio lower than 10 and 
producing primary \chem{N}{14} (Frost 1997; FC97). This happens in the
more massive AGB stars.
At very high T$_{cb}$, other reactions begin to occur (NeNa and MgAl chains); 
in particular, in extreme cases, \chem{Al}{26} can possibly be directly 
produced inside the convective envelope (FC97). 
This nuclide is particularly interesting for what concerns $\gamma$-ray 
line astronomy (Prantzos 1996) and cosmochemistry (Anders \& Zinner 1993).

\subsection{Mass loss, 3Dup, HBB and Transport Processes}
As we just discussed, AGB stars undergo a very rich and unique 
nucleosynthesis. 
Recurrent occurences of the 3Dup enrich the stellar surface with freshly 
synthesized nuclides which are then ejected into the ISM through the 
strong winds. 
On the other hand, for stars with masses higher than 4\,M$_{\odot}$, 
the 3Dup signatures may be modified by HBB.

There is actually a very complex interplay between mass loss, 3Dup and HBB.
Mass loss affects the AGB duration, and consequently the number of pulses 
and subsequent 3Dup events.
In addition, a minimum envelope mass is required for HBB and 3Dup to occur. 
As the envelope mass decreases, HBB may be shutdown long before the 3Dup
ends. The surface composition in the latter stages of the evolution
thus depends critically of the competition between these effects (Frost
1997). 
This is of crucial importance for example for the formation of super-lithium 
rich stars or of high luminosity C stars, as found in the 
Magellanic Clouds (van Loon et al. 1997). It was shown for example that, for
peculiar mass loss rates, high luminosity C stars can be obtained (Frost
et al. 1998); this casts some doubt on the previously expected maximum 
luminosity for these objects. 

Unfortunatly, very high uncertainties concerning both the 3Dup mechanism
and mass loss remain. 
The 3Dup is difficult to obtain numerically (see Lattanzio 1989). 
Its depth depends on the assumptions made at the convective region 
boundary (Frost \& Lattanzio 1996), as well as on the way mixing is 
handled within the calculations (Straniero et al. 1997; Frost 1997). 

On the other hand, the way following which the mass loss rate increases 
with time along the TP-AGB phase is still questionable (Bryan et al. 1990;
Vassiliadis \& Wood 1993; Bl\"ocker 1995). 
Since  the various nucleosynthesis processes are highly sensitive to 
the adopted mass loss rate, this results in important uncertainties on the 
evolution of the surface composition up to the AGB tip. 

\subsection{Some Observational Constraints for Future Developments}
Some important failures of the present models should help us to identify 
some processes which are not yet included in the computations. 
Lets us just recall the most striking points. 

As mentioned is \S 2.2, the amount of \chem{C}{13} available for the
s-process nucleosynthesis is by far insufficient to explain the observed
distribution of s-elements in the solar system. 
The additional \chem{C}{13} source needed has not be identified. 
The most natural way to conciliate the theoretical models with the 
observations would be to increase the amount of \chem{C}{13} in the 
intershell region. A slow particle process able to transport protons 
from the HBS downwards during the interpulse phase could substantially
help to enhance both the classical and radiative s-processes. 

Predictions for the formation of super-lithium rich stars (Sackmann \& 
Boo\-throyd 1992) reproduce well the observations in the 
Magellanic Clouds (Smith \& Lambert 1989, 1990b; Plez et al. 1993; Smith
et al. 1995). However, the rather high level of Li production detected at 
the surface of some evolved but relatively faint AGB stars of our Galaxy 
still remains unaccountable by the present models. 
Indeed, these galactic objects, which are also carbon stars, have 
relatively low initial mass, and should not undergo HBB.  
The explanation could also come from the radiative slow transport 
process mentioned above. The inclusion of such a mechanism may have 
an important impact on the production of \chem{Li}{7} in the Galaxy. 

The current models fail to reproduce the \chem{O}{16}/\chem{O}{17} 
and \chem{O}{16}/\chem{O}{18} ratios obtained from the analysis of 
interstellar oxide grains (Nittler et al. 1994) and from  
observations of low mass S and C AGB stars (Harris et al. 1985, 1987;
Kahane et al. 1992). Boothroyd et al. (1995) suggested that to conciliate
the data with the observations, one probably has to invoke a slow 
particle transport referred to as ``cool bottom processing". 

Las but not least, some carbon stars are observed, especially in the 
galactic bulge, with low luminosity at which the models fail at 
producing the occurrence of the 3Dup. 
Solutions to increase the depth of the convective envelope during a Dup, e.g.
by treating in some way semi-convection (Lattanzio 1986) or by
invoking undershooting (Alongi et al. 1991) have been suggested. They
indeed help the 3DUP to occur but do not yet solve completely the problem.
Fundamentally, this could be due to our bad knowledge of convection inside
stars. 

\section{Yields}
Numerous observations indicate that ``non-standard" particle transport 
processes are acting in LMS at different phases of their evolution, that
may substancially modify our predictions concerning the contribution of 
these objects to the chemical evolution of galaxies (see for example the
\chem{He}{3} problem). 
In addition, in LIMS, various nucleosynthesis processes are 
highly sensitive to the mass loss and the 3Dup, which are poorly known. 

Despite the remaining uncertainties, one can conclude saying that LIMS 
are rather significant producers of \chem{He}{4}, \chem{Li}{7}, 
\chem{C}{13}, \chem{N}{14} (see Fig.1; HBB in AGB stars may be an important 
source of primary \chem{N}{14} that has to be taken into account in the 
chemical evolution models), \chem{O}{17}, \chem{F}{19}, \chem{Ne}{22},
\chem{Na}{23}, \chem{Mg}{25} and \chem{Al}{26}. 
They are also responsible for the main component of the solar system 
heavy element distribution. 
They partially deplete the ISM content in \chem{N}{15}, \chem{O}{16}, 
\chem{O}{18}, \chem{Ne}{20}, \chem{Mg}{24} and  \chem{Si}{28}.

Future insight in the physics of LIMS (extra-mixing processes during the RGB 
and AGB phases, additional source of \chem{C}{13}, 3Dup, wind mechanism,
history and rates of mass loss, ejection of the planetary nebulae) 
should improve, both qualitatively and quantitativaly, the predictions 
for the final yields of these objects.

\begin{figure}[ht]
\plotfiddle{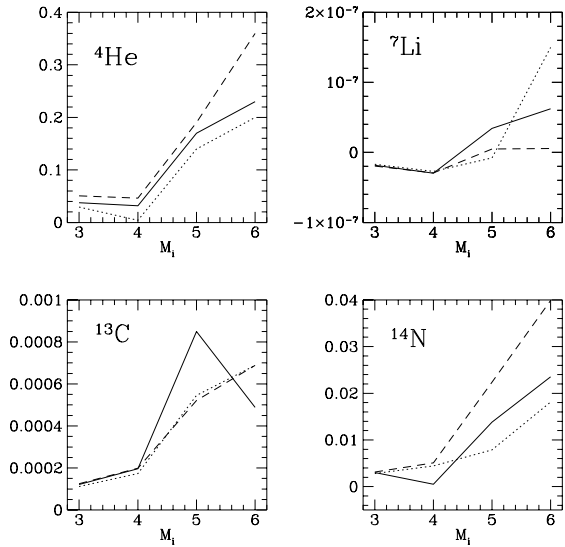}{65truemm}{0}{40}{40}{-160}{-80}
\caption{Net contribution, i.e. total final mass (in M$_{\odot}$) ejected 
reduced by the amount of matter that would have been ejected with the initial
composition (here solar), for a few elements as a function of the initial 
stellar mass. Predictions are given only for IMS (FC97), since missing
processes preclude precise results for LMS. The solid lines refer to
``standard" mass loss rates, the dotted lines to mass loss rates
increased or decreased by a factor of two}
\end{figure}

\bigskip 
\bigskip 
\acknowledgments
I express my thanks to Manuel Forestini for a careful reading of the 
manuscript.

\end{document}